\documentclass[conference]{IEEEtran}
\IEEEoverridecommandlockouts
\usepackage{cite}
\usepackage{amsmath,amssymb,amsfonts}
\usepackage{algpseudocode}
\usepackage{algorithm}
\usepackage{mathrsfs}
\usepackage{graphicx}
\usepackage{textcomp}
\usepackage{xcolor}
\usepackage{xurl}
\usepackage{booktabs, tabularx}
\restylefloat{table}
\usepackage{multirow}
\usepackage{makecell}
\usepackage{adjustbox}

\usepackage{eso-pic}
\newcommand\AtPageUpperMyright[1]{\AtPageUpperLeft{%
 \put(\LenToUnit{0.17\paperwidth},\LenToUnit{-2cm}){%
     \parbox{0.9\textwidth}{\raggedleft\fontsize{8}{11}\selectfont #1}}%
 }}%
\newcommand{\conf}[1]{%
\AddToShipoutPictureBG*{%
\AtPageUpperMyright{#1}
}
}    

\newcommand\omicron{o}
\newcommand{\ceil}[1]{\left\lceil #1 \right\rceil}
\newcommand{\floor}[1]{\left\lfloor #1 \right\rfloor}

\newcommand{\probP}{\text{I\kern-0.15em P}}
\newcommand{\ie}{\textit{i.e.}}

\def\BibTeX{{\rm B\kern-.05em{\sc i\kern-.025em b}\kern-.08em
    T\kern-.1667em\lower.7ex\hbox{E}\kern-.125emX}}
\begin{document}

\title{\vspace*{1cm} Develop End-to-End Anomaly Detection System
\thanks{This work is done during the internship of Emanuele Mengoli at Cisco Meraki.}
}

\author{\IEEEauthorblockN{Emanuele Mengoli}
\IEEEauthorblockA{\textit{\'Ecole Polytechnique} \\
Paris, France \\
em.mengoli@gmail.com}
\and
\IEEEauthorblockN{Zhiyuan Yao}
\IEEEauthorblockA{\textit{Cisco Meraki} \\
Paris, France \\
yzhiyuan@cisco.com}
\and
\IEEEauthorblockN{Wutao Wei}
\IEEEauthorblockA{\textit{Cisco Meraki} \\
San Francisco, USA \\
wutwei@cisco.com}
}

\maketitle
\conf{\textit{Proc. of the International Workshop on User Understanding from Big Data Workshop (DMU2 - IEEE ICDM 2023) \\ 
1-4 December 2023, Shanghai, China}}

\begin{abstract}
Anomaly detection plays a crucial role in ensuring network robustness. 
However, implementing intelligent alerting systems becomes a challenge when considering scenarios in which anomalies can be caused by both malicious and non-malicious events, leading to the difficulty of determining anomaly patterns.
The lack of labeled data in the computer networking domain further exacerbates this issue, impeding the development of robust models capable of handling real-world scenarios.
To address this challenge, in this paper, we propose an end-to-end anomaly detection model development pipeline.
This framework makes it possible to consume user feedback and enable continuous user-centric model performance evaluation and optimization.
We demonstrate the efficacy of the framework by way of introducing and bench-marking a new forecasting model -- named \emph{Lachesis} -- on a real-world networking problem.
Experiments have demonstrated the robustness and effectiveness of the two proposed versions of \emph{Lachesis} compared with other models proposed in the literature.
Our findings underscore the potential for improving the performance of data-driven products over their life cycles through a harmonized integration of user feedback and iterative development.
\end{abstract}

    
\section{\uppercase{Introduction}}
\label{sec:introduction}

Anomaly detection systems are a category of purpose-built techniques designed to predict and prevent abnormal occurrences within complex systems through continuous real-time monitoring~\cite{laptev2015generic,vallis2014novel}.
These systems have the ability to anticipate upcoming data values, thereby facilitating informed decision making. In the context of networking, the indispensability of such systems is underscored by their pivotal role in performing network health management functions, thereby ensuring seamless operational continuity~\cite{cisco_v1}.

The focus of these models lies in the domain of prediction execution, which involves determining an appropriate threshold for triggering alerts~\cite{survey_2013_ml_techniques}.
This paper is situated within this domain and focuses its attention on the issue of anomalies related to system performance.
Our endeavour extends the groundwork established in~\cite{cisco_v1}, with a particular emphasis on the use of supervised techniques for the detection and identification of anomalies in the computer networking domain.

It is noteworthy that these systems often lack to identify the root cause of anomalous events.
While identifying and quantifying responsibility for mono-causal events can be a straightforward exercise, the complexity of the task increases significantly when faced with scenarios involving multiple causal factors.
The identification of anomaly causation is of great interest to users and network operators.
In particular, supervised methods require the availability of anomaly ground truth data to establish causal responsibilities.

In this paper, we present a modularized framework that allows continuously evaluating and optimizing a supervised anomaly detection model.
The foundation and the demonstration of our framework is anchored in the context of predicting abnormally high amount of networking event occurrences in networking switches developed by Cisco Meraki.

The development of an anomaly detection system with the ability to continuously quantify the responsibility of potential root causes presents a number of challenges.

\textbf{Challenge 1: Extensible Labelled Data Acquisition.}
The construction and fine-tuning of anomaly detection systems are fundamentally grounded in the availability of a substantial and well-labeled dataset.
However, achieving this foundation is met with intrinsic challenges owing to the multifaceted nature of networking systems.
We tackle this challenge by introducing an adaptive and extensible predictive model that can evolve in tandem with the network it monitors, thereby sustaining a relevant and acute predictive capability through the incorporation of user feedback loops.

\textbf{Challenge 2: Accuracy - Responsiveness trade-off.}
There is a delicate balance between accuracy and real-time responsiveness.
Achieving high predictive accuracy is an ultimate goal in developing anomaly detection systems, yet the increased latency introduced by substantial models impairs the timely identification and mitigation of anomalies.
Our proposed algorithm achieves a synthesis of these attributes by combining two proposed versions of the time-series models.
The first prioritises accuracy in the prediction phase, followed by a second version adapted to real-time implementation.
The latter is designed to adapt predictions based on the degree of data deviation from the training time series.

\textbf{Challenge 3: Scalability.}
The expansive and ever-growing network infrastructures dictate a prerequisite for the anomaly detection system to exhibit a high degree of scalability.
Effective deployment requires that the models are able to deliver tailored performance with inherent flexibility.
Maintaining scalability should be intrinsic to both the initial development and the ongoing maintenance of the system, ensuring not only a reactive adaptability to current network configurations but also a proactive readiness to accommodate future growth and complexities with minimal disruptions.

The structure of the paper is as follows:
Section~\ref{related_work} provides a literature review over related anomaly detection systems.
Section~\ref{problem_formulation} describes the essential background information on the problem in which we demonstrate our proposed solution, and formulates the problem to solve.
The designed framework and the proposed algorithm is explained in Section \ref{methodology}.
Subsections \ref{lachesis_v1} to \ref{lachesis_v2} detail specific components of the proposed anomaly detection model.
The evaluation process is described in Section \ref{evaluation}.
Finally, concluding remarks are presented in Section \ref{conclusion}.

\section{Related work} 
\label{related_work}

The literature landscape revealed a substantial body of work on anomaly detection, focusing mainly on network intrusions rather than performance anomalies.
However, common tools are used to detect anomalous behaviours, namely supervised and unsupervised techniques, so a literature review was conducted on these techniques.

Unsupervised techniques demonstrated the ability to effectively detect anomalous patterns.
However, their sensitivity to significant fluctuations in the data made it difficult to distinguish between normal and anomalous patterns~\cite{bhuyan2013network}.
Empirical evidence highlighted the superior performance of supervised learning over unsupervised methods, provided that the test data did not contain unknown attacks~\cite{survey_2013_ml_techniques}.
In the context of real-time implementation, supervised techniques were preferred due to their less resource-intensive testing phase~\cite{bhuyan2013network}.

As shown in~\cite{ericsson_anomaly}, Ericsson Lab has proposed a customised analytical engine for anomaly detection in mobile networks. This solution employs an Autoregressive Integrated Moving Average (ARIMA) model to predict Key Performance Indicator (KPI) values and a heuristic algorithm to calculate the actual value as ground truth. This approach is particularly suitable for scenarios with low data volatility; however, in cases of greater fluctuation, reliance on labelled data becomes essential to accurately distinguish normal from abnormal behaviour and to avoid estimation errors.

An unsupervised algorithm based on Isolation Forest (iForest) has been proposed for anomaly classification in~\cite{network_anomaly_iforest}.
While this approach showed robustness under the assumption of stationarity or periodicity of the time series, it may lack precision in the presence of high data volatility. 
In such cases, labelled truth data are essential to build accurate anomaly models.

Furthermore, \cite{stad_anomaly} introduced an unsupervised spatio-temporal anomaly detector consisting of two stages: geographical anomaly identification and temporal anomaly detection.
This system excelled in detecting anomalous behaviour related to geographical location and time, thus addressing the need for timely intervention to maintain network resilience and service availability.
Using One Class SVM (OCSVM), Long Short-Term Memory (LSTM) and Support Vector Regression (SVR) recurrent neural networks, this approach outperformed iForest and ARIMA in spatial and temporal detection, respectively.

Authors in~\cite{causality_semi_supervised_rf} proposed a model that requires a set of previously labelled data to train the semi-supervised root cause analysis, leading to a classification problem centred on matching the root cause, using a random forest classifier.
Pre-labelled data may be complex to obtain since it requires full knowledge of the system dynamics. This could be extremely challenging to achieve, especially in a scenario that is characterized by multiple causes, for which specific responsibility is hard to define or quantify, as in our scenario.

The authors of~\cite{causality_unsupervised_fsm} proposed an unsupervised method of inferring root causes in mobile networks using finite state machines (FSM) based on Principal Component Analysis (PCA), which reduces dimensionality and extracts meaningful data from the original dataset.
FSM identifies problematic procedures that result in cascading failures in subsequent procedures.
The identification of the specific problem is made possible by translating the error codes of the procedures once the original anomalous pattern is identified.
The results of FSM are intriguing when applied to a time series or dataset with moderate-sized normal and anomalous patterns, such as a specific messages sequence shown in~\cite{causality_unsupervised_fsm}.
However, this method is combinatorial in nature.
As a result, the complexity is likely to increase considerably when considering the stochastic nature of normal and abnormal patterns.

In~\cite{causality_unsupervised_transferL}, the authors proposed an unsupervised model built in two-stage clustering and further improved by applying transfer learning to historical data.
This system requires a deterministic pass/fail classifier to detect anomalous behaviour in the data samples.
Achieving this can be difficult if normal patterns follow stochastic behaviour.

An autonomous diagnostic system within a self-healing network is proposed in~\cite{causality_lte_unsupervised}.
It uses a self-organising map (SOM) to generate clusters.
Expert-based labelling is performed on the clusters, guided by their statistical characteristics.

Unlike prior studies, this paper proposes a simple yet effective model development pipeline that allows continuously develop and optimize user-centric anomaly detecting systems in an extensible manner.
Given the focus of this paper on high-fluctuation data exhibiting pattern similarities between malicious and non-malicious causes, it is preferred to use a supervised method supported by labelled ground truth data.
Furthermore, the unique statistical and temporal properties associated with each node introduced challenges to classical supervised approaches. In light of these considerations, we had introduced \emph{
Lachesis}, a particle-filter-based anomaly detection algorithm. To the authors knowledge no particle-filter-based anomaly detection system for network management has been proposed in the literature to date.
To assess the forecasting accuracy and performance of our proposed model, we had conducted a comparative analysis with the following algorithms: Prophet~\cite{prophet,prophet_pkg}, statistical model~\cite{cisco_v1} (named phase $1$ for the purpose of this paper), Dynamic Linear Model (DLM)~\cite{dlm, dlm_pkg}, Autoregressive Integrated Moving Average (ARIMA)~\cite{Jiang_KATS_2022,arima}, Seasonal Autoregressive Integrated Moving Average (SARIMA)~\cite{Jiang_KATS_2022,sarima}, linear and quadratic regression~\cite{sck_linear_regression,sck_quad_regression}.

\begin{figure*}[t] 
    \includegraphics[width=\textwidth]{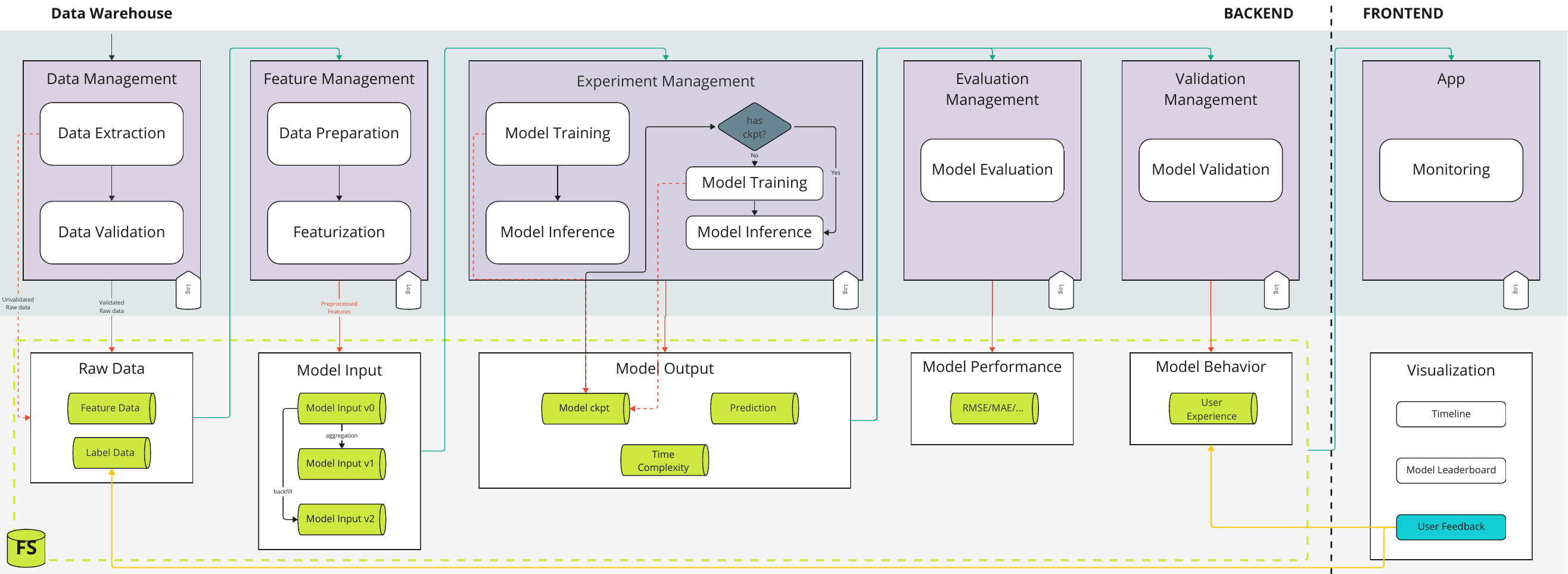}
    \caption{Model development lifecycle taking user feedback into the loop.}
    \label{fig:dev-loop}
    \vskip -.1in
\end{figure*}

\section{Problem Formulation}
\label{problem_formulation}

In this paper, we exemplifies the model development pipeline for anomaly detection system development in the context of a typical event in networking switches.
In particular, we focus on Layer-$2$ address flapping detection, commonly referred as Media Access Control (MAC) flapping detection.
For consistency with the nomenclature, we will use the term MAC-flap detection when referring to this type of event.

\begin{figure} [!t] 
\includegraphics[width=\columnwidth]{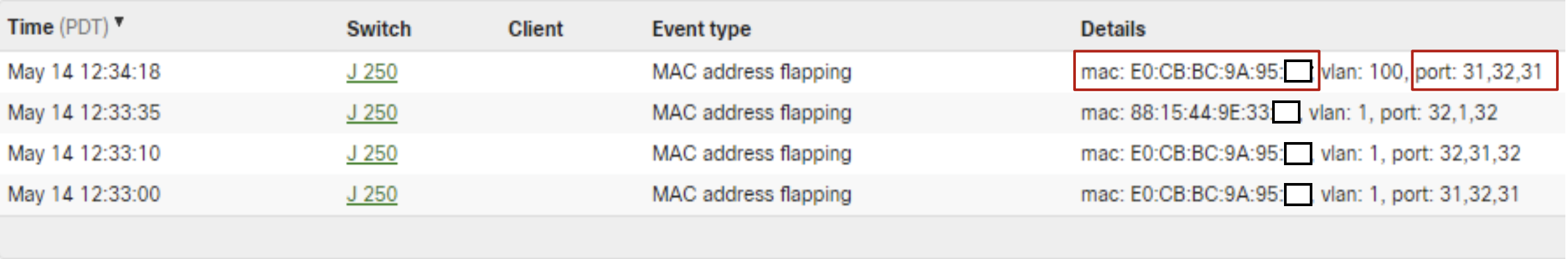}
\caption{MAC-flap detection.}
\label{fig:MAC-flap_detection}
\vskip -.1in
\end{figure}

MAC-flap event occurs when a MAC address is learned $3$ times or more on $2$ or more different ports within $10$ seconds on a networking switch~\cite{macflap_loop_detected}. 
MAC-flap events can cause packet loss in the network.
Meraki switches include MAC-flap detection as a standard feature, which monitors the MAC forwarding table and reports flap events on the dashboard, as shown in Figure~\ref{fig:MAC-flap_detection}.
However, switches do not have a built-in intelligence mechanism that triggers detection only in response to real problems.
As a result, this alerting system can be overwhelmed by both true and false positives, generating noise to the end users.
At the mean time, the root causes of MAC-flapping are not intuitive.
This problem particularly affects the troubleshooting activity of network administrators, as the burst visibility affects the identification of the root cause of anomalies.

The goal of building an intelligent alert system is to help identify a baseline for each networking switch and trigger an alert only when the amount of MAC-flap events surpasses the derived baseline.
This process can be abstracted as a time-series forecasting problem, \ie, given the past $5$ weeks' historical data of the number of MAC-flap events, we want to predict the next week's MAC-flap events for each node.
Then the second step is to identify whether the generated alerts correspond to actual networking issues in the network, based on which we can evaluate the alert accuracy.

\section{Methodology}
\label{methodology}

To tackle this real-world networking problem and its corresponding challenges, we propose a model development framework, accompanied by a particle-filter-based alerting system (named \emph{Lachesis}).
They are orchestrated to minimise false positive alerts, allowing further investigation of potential root causes and to determine the share of responsibility.
Lachesis learns from the historical data to generate a baseline. 
It forecasts future data values and predicts an upper-bound on the number of MAC-flap events in a given time window, above which an alarm is triggered.
The model development pipeline then allows to iteratively and continuously benchmarking and optimizing the alerting systems using various data-driven methods.

\subsection{Model Development Pipeline}
\label{mlops}

An automated pipeline is a cornerstone in orchestrating scalable, extensible, reproducible, and efficient model development.
We propose a model development pipeline depicted in Figure~\ref{fig:dev-loop}, which not only lays down a robust blueprint for ML operations but transcends into a facilitator for a refined user experience through the incorporation of user feedback.

User feedback emerges as a cardinal element, steering the evolvement of the system through enriched labeled data procurement and user experience enhancement.
Networking systems are characterized by dynamic state transitions that not only occur at large scales but also transpire rapidly, often outpacing the ability of existing systems to identify and tag them appropriately.
Moreover, the underlying triggers of anomalous events are diversified, emanating from a range of factors including diverse protocol implementations and their respective versions.
At the mean time, when developing data-driven models, it is important to take user experience into account and align user expectation and requirements with our data-driven products.
Our proposed pipeline is able to resolve this by incrementally absorbing user feedback in an extensible manner.

The proposed pipeline (Figure~\ref{fig:dev-loop}) comprises $6$ components unified to engender a data-driven solution for forecasting and generating alerts pertinent to the anomaly detection system outlined in this paper.
Each component is designed to bring a seamless transition from data procurement to user feedback integration, anchoring on a symbiotic relationship between the system and its users to facilitate continuous improvement through iterative learning and feedback assimilation.

\subsubsection{Data Management}

This component is responsible for extracting and validating data obtained from diverse sources.
These processes are orchestrated in a structured data warehouse.
We expound on the measures adopted to validate the extracted data, fostering a repository enriched with credible and quality data, which forms the bedrock for feature management and model training.

\subsubsection{Feature Management}

This component is responsible for preparing the raw data into a format ready for feature extraction, as well as the transformation of prepared data into features that are conducive for the model learning process.
Making this component independent helps control the versions of input data for various models, meanwhile, facilitate feature selection and extraction.

\subsubsection{Experiment Management}

On the heart of the pipeline, this component is responsible for model training and inference.
It offers the flexibility in accommodating iterative learning through checkpoints and retraining, ensuring a progressive model evolution.
Besides model checkpoints and predictions generated by the models, we also track and evaluate the computational complexity of the model.

\subsubsection{Evaluation Management}

This component is separated from the experiments managed above so that jobs can be scheduled in parallel.
Decoupling the evaluation also allows for seamlessly updating evaluation metrics without re-running heavy jobs of experiments (model training and inferring).
In the context of MAC-flap alerting systems, MAE, MSE, and RMSE are utilized in gauging the model's forecasting prowess.

\subsubsection{Validation Management}

This key component bridges the technical finesse of the model and the practical business requirements and user experiences.
Besides the technical evaluation mentioned above, it is imperative to ensure that the developed model aligns well with the broader business objectives.
The most significant objective of customer-facing data-driven products is dedicated to analyzing and integrating user feedback to create a model that is responsive to the user's requirements and preferences.
We create a channel for users feedback on various facets such as false positives/negatives, alert sensitivities, and other experiential aspects. 
By assimilating this direct feedback from end-users, we enrich the validation process, improving user satisfaction metrics and operational efficiency.

\begin{table}[t]
    \footnotesize
    \centering
    \caption{Evaluation and Validation Metrics}
    \label{tab:evaluation_metrics}
    \begin{tabular}{p{0.38\linewidth} p{0.55\linewidth}}
        \toprule
        \textbf{Metric} & \textbf{Description} \\
        \midrule
        \multicolumn{2}{l}{\textbf{Evaluation (Regression)}} \\
        \midrule
        MSE & Mean Squared Error \\
        \addlinespace
        RMSE & Root Mean Squared Error \\
        \addlinespace
        MAE & Mean Absolute Error \\
        \addlinespace
        \midrule
        \multicolumn{2}{l}{\textbf{Computational Complexity}} \\
        \midrule
        $t_{\text{train}}$ (s) / 1k data points & Time for training per 1k data points \\
        \addlinespace
        $t_{\text{infer}}$ (s) / 1k data points & Time for inference per 1k data points \\
        \addlinespace
        $t_{\text{train}}$ (s) / 1k nodes & Time for training per 1k nodes \\
        \addlinespace
        $t_{\text{infer}}$ (s) / 1k nodes & Time for inference per 1k nodes \\
        \addlinespace
        \midrule
        \multicolumn{2}{l}{\textbf{Validation (User Experience)}} \\
        \midrule
        Avg. Daily Alerted Nodes & Average
        number of nodes alerted per day\\
        \addlinespace
        Avg. Daily Alerts & Average number of alerts per day\\
        \addlinespace
        Total Alerts & Total alerts in whole prediction time-span\\
        \addlinespace
        Avg. Alert Duration (min.) & Average alert duration\\
        \addlinespace
        Std. Alert Duration (min.) & Standard deviation alert duration\\
        \addlinespace
        Avg. Alerts per hour & Average alerts per hour\\
        \addlinespace
        Std. Alerts per hour & Standard deviation alerts per hour\\
        \addlinespace
        Accuracy & $\frac{TP + TN}{\text{Total}}$ \\
        \addlinespace
        Precision & $\frac{TP}{TP + FP}$ \\
        \addlinespace
        Recall & $\frac{TP}{TP + FN}$ \\
        \addlinespace
        Specificity & $\frac{TN}{TN + FP}$ \\
        \addlinespace
        Balanced Accuracy & $\frac{\text{Recall} + \text{Specificity}}{2}$ \\
        \bottomrule
    \end{tabular}

\end{table}

\subsubsection{Frontend Application}

The frontend application is the user-facing component of our framework, designed to be an interactive platform where users can engage directly with the system.
It gives insights into the various performance indicators and presents a graphical representation of the forecasting timelines, allowing users and stakeholders to visualize the efficacy of the alerting system over periods.
The frontend application provides a two-way communication channel, allowing users to provide feedback on the forecasting and alert generation process directly.
Moreover, this interface enables continuous learning, as it facilitates the collection of labeled data through user feedback, thereby integrating an incremental improvement in the model's accuracy. 

\begin{figure}[t]
    \centering
    \includegraphics[width=.75\columnwidth]{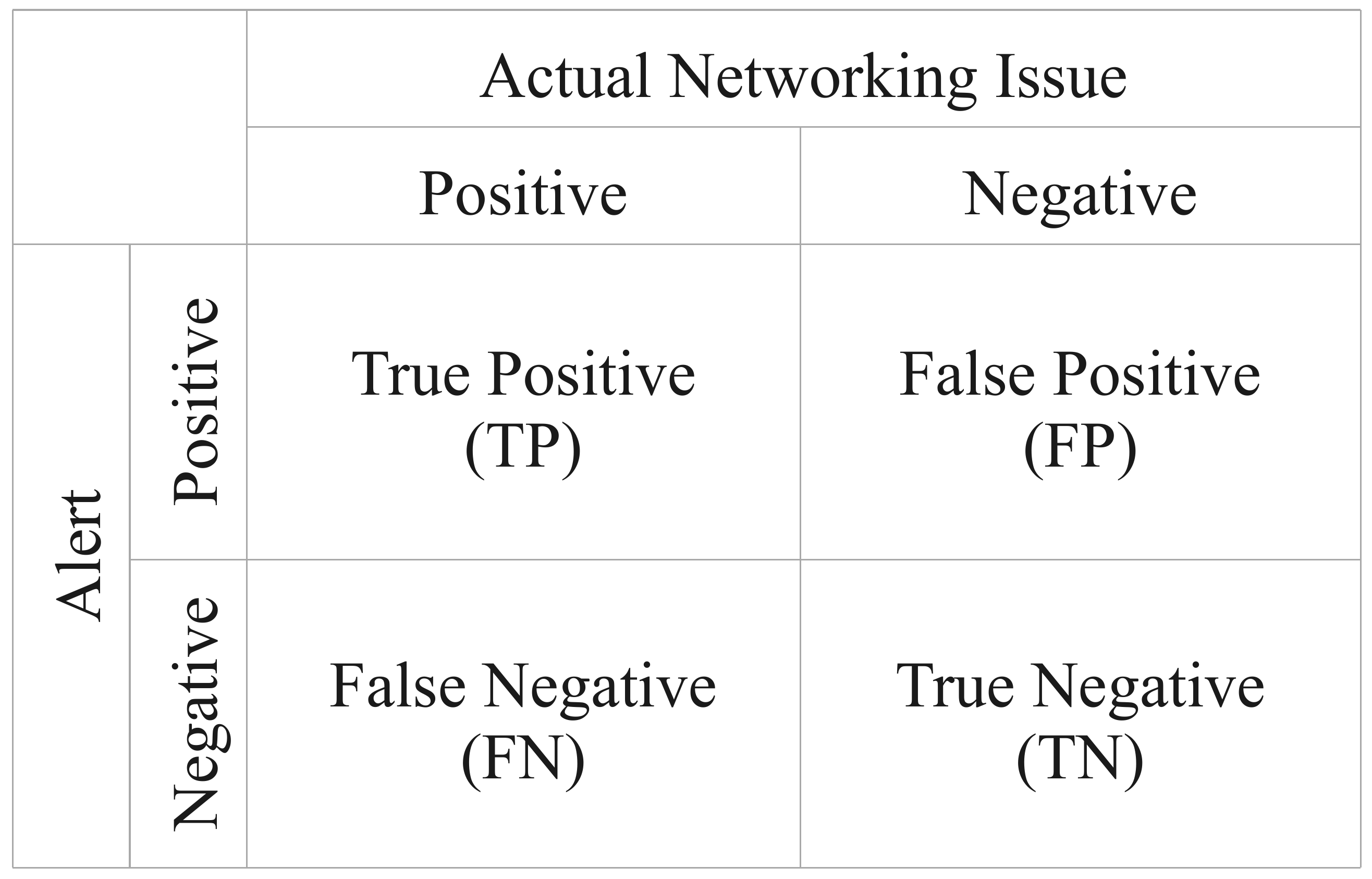}
    \caption{Confusion matrix of MAC-flap alerts.}
    \label{fig:classification}
    \vskip -.1in
\end{figure}

Whilst the potential root causes of MAC-flap events vary across different networking setups, this paper focuses on one root cause identified via this model development pipeline -- switch network loop~\cite{macflap_loop_detected}.
Network loops occur when a switch detects its own MAC address on a received loop detection control packet~\cite{macflap_loop_detected}.
This can be avoided by activating the \textit{Spanning Tree Protocol (STP)} function which determines the active ports among the switches of the topology~\cite{stp}. 
The whole list of evaluation and validation metrics covered by the pipeline throughout all the components is summarized in Table~\ref{tab:evaluation_metrics}.
Model evaluation is conducted with $3$ regression metrics.
Computational complexity is assessed over both the number of time-series data points and number of nodes (network switches).
Model validation is based on user experience.
We measure the alert frequencies and ratio so as to consolidate whether the alerts become noisy for the end-users.
Besides, we assess the accuracy of the alerts based on the confusion matrix depicted in Figure~\ref{fig:classification}, \textit{e.g.} if an alert is raised for a node within $1$ hour difference to when an actual networking issue (\ie a network loop is detected), the alert is classified as a true positive.

\subsection{Lachesis Model}
\label{lachesis_general}

To tackle the time-series forecasting problem for MAC-flap events, $2$ versions of the predicting model Lachesis are proposed and assessed with the proposed pipeline.
The underlying intuition of Lachesis is that anomalies of MAC-flap events can result from non-malicious causes, such as client roaming across multiple access points, recurring consistently on the same day of the week and at the same time.
Conversely, it links malicious anomalies to sporadic events occurrences. 
Predictions are made for each node given the time series data of timestamps and respective number of MAC-flap events, from which weekday \textit{dw} and time \textit{t} are then extrapolated.

\subsection{Lachesis version 0}
\label{lachesis_v1}

The first version of the Lachesis model takes five weeks of historical data, and then makes predictions for the sixth week.

The user must specify the time aggregation parameter and the granularity in order to divide the series into a succession of discrete signals in the time domain. 

In our case, a time aggregation of $1$ hour is considered, with granularity to the minute.
The user sets the forecast flag $\eta$, deciding whether to run a forecast or an upper-bound estimation.

The time of day is used to set the beginning of the time bucket obtained by grouping the time series on a daily basis and according to the time aggregation parameter $\tau$.
After grouping the input time series by date and time of day, each signal is transformed into the frequency domain using the Fast Fourier Transform (FFT) \cite{fft}, resulting in a discrete frequency transform of the data $\mathrm{Z}_k$.
The frequency domain is used both to reduce the size of the dataset and to improve the visibility of recursive values.

By grouping these values by day of the week and time slot $(dw,t_\tau)$, a significant value is obtained for each frequency $\kappa$ through the application of a custom lambda function:

\begin{align}
    \tilde{\mathrm{Z}}_\kappa &= \lambda(\mathrm{Z}_\kappa)^{dw,t_\tau} \notag =\\
    &= \left[ \left(\sqrt{\hat{\mathrm{Z}}_k} \cdot \lvert \Bar{\mathrm{Z}}_k + 3\sigma \rvert \right) + \sqrt{\operatorname{card}(\mathrm{Z}_k)} \right] \cdot \mathrm{c} \notag \\
    &\forall dw, t_\tau \in g_\tau
\end{align}
 
In which $\hat{\mathrm{Z}}_k$ is the maximum value of $\mathrm{Z}_k$, $\Bar{\mathrm{Z}}_k$ the mean, $\sigma$ the standard deviation, $\operatorname{card}(\cdot)$ the cardinal operator and $g_\tau$ the original time series bucketed by $\tau$. $\mathrm{c}$ is a coefficient that assumes significance in upper-bound estimation, being set at $1.5$. This value was determined through empirical testing, that is, the value that minimized the RMSE metric, given the range $[1.1, 2]$. 

For each date-time bucket $(d,t_\tau)$, the Power Spectral Density (PSD) $S_{zz}$ is calculated. Our goal is to highlight the set of frequencies that most contribute to the $S_{zz}$ peaks, so we compute the covariance matrix $\Sigma_{\tilde{\mathrm{Z}}_\kappa, S_{zz}}$ and compute its eigenvalues and eigenvectors. By scaling the eigenvectors by the eigenvalues, we obtain a vector subspace $\mathrm{W}(\mathrm{v})$ indicating the main direction and magnitude of the contribution of $\tilde{\mathrm{Z}}_\kappa$ to the $S_{zz}$.
After calculating the centre $\overline{\mathrm{W}}(\mathrm{v})$ of the vector subspace, we perform Density-Based Spatial Clustering of Applications with Noise (DBSCAN) on $\tilde{\mathrm{Z}}_\kappa$ to find the subset of $\tilde{\mathrm{Z}}_\kappa$ whose centroid is closest to $\overline{\mathrm{W}}(\mathrm{v})$. The normalisation of $\tilde{\mathrm{Z}}_\kappa$ to $\mathrm{W}(\mathrm{v})$ is invariant due to the linearity of the transformation, so we keep the original value. On the selected cluster, we perform Kernel Density Estimation (KDE) to obtain an estimate of the probability density function of the anti-transformed values. The obtained density function is injected into the inspired particle filter to evaluate the posterior distribution of the particles in the time domain. particles are generated using a uniform distribution in the range defined by the minimum and maximum of the selected cluster.
From the newly generated set of particles, those whose probability is above the user defined probability threshold are retained. 
For accuracy, the threshold should be specified in the neighborhood of $1$, in order to select particles with high probability of being a correct prediction.

From these, the average is calculated as a result of the number of MAC-flap events expected for that time slot.

The details of the algorithm are presented in Algorithm \ref{alg:lachesis_v1}.

\begin{algorithm}
    \footnotesize
    \caption{Lachesis version 0}\label{alg:lachesis_v1}
    \begin{algorithmic}
    \Require $dw(d,t)$ \Comment{Compute the weekday $dw$ from the date $d$ and time $t$}
    \Require $g(d,t)$ \Comment{Compute the MAC-flap event numbers at date $d$ and time $t$}
    \Require $\tau$ \Comment{Time aggregation parameter, minute-based}
    \Require $\eta$ \Comment{Forecast flag}
    \Require $\epsilon$ \Comment{Particle Filter probably threshold}
    \Require $\theta$ \Comment{Number of particles}
    \Require $ke$,$bw$ \Comment{KDE kernel function, kernel bandwidth}
    \Require $\epsilon'$,$ms$ \Comment{DBSCAN Max. distance between samples, Min. number of samples}
    \Ensure $g(d,t) \geq 0$
    \State $g_\tau \gets \{g(d,t)\}_\tau$ 
    \Comment{Sequence of $g(d,t)$ bucketed by $\tau$}
    \State $N \gets 35$ \Comment{Time window size in days}
    \State $Y \gets \left[\right]$ \Comment{Result}
    \If{$\eta$ is True}
        \State $\mathrm{c} \gets 1$ \Comment{Forecast}
    \Else
        \State $\mathrm{c} \gets 1.5$ \Comment{Threshold prediction}
    \EndIf
    \While{$N > 0$}
        \State $\mathrm{Z}_\kappa \gets \mathscr{F}[g_\tau]$ \Comment{FFT}
        \State $N \gets N-1$ 
    \EndWhile
        \State $\tilde{\mathrm{Z}}_\kappa \gets \left[ \left(\sqrt{\hat{\mathrm{Z}}_k} \cdot \lvert \Bar{\mathrm{Z}}_k + 3\sigma \rvert \right) + \sqrt{\operatorname{card}(\mathrm{Z}_k)} \right] \cdot \mathrm{c}$
        \State $ \qquad \qquad \qquad \qquad \qquad \qquad \forall dw, t_\tau \in g_\tau$
    \State $W \gets 7$ \Comment{Prediction timespan in days}
    \While{$W> 0$}
        \State $T \gets \left\{t_\tau \right\}_\tau$ 
        \Comment{Time slot sequence}
        \While{$\lvert T \rvert > 0$} \Comment{Sequence cardinality}
    
            \State $S_{zz} \gets \operatorname{PSD}(\tilde{\mathrm{Z}}_\kappa)$ 
            \State $\Sigma_{\tilde{\mathrm{Z}}_\kappa, S_{zz}} \gets \operatorname{cov}(\tilde{\mathrm{Z}}_\kappa,S_{zz})$ 
            \State $\zeta, \mathrm{v} \gets \operatorname{eigen}(\Sigma_{\tilde{\mathrm{Z}}_\kappa, S_{zz}})$
            \State $\mathrm{W}(\mathrm{v}) \gets \zeta \times \mathrm{v}$ 
            \State $\mathrm{C}_\kappa \gets \operatorname{DBSCAN}(\tilde{\mathrm{Z}}_\kappa, \epsilon', ms)$ 
            \State $\tilde{\mathrm{C}}_\kappa \gets \arg\min(\lvert \lvert \mathrm{C}_\kappa - \overline{\mathrm{W}}(\mathrm{v})\rvert \rvert_2)$ 
            \State $\Upsilon_{{\tilde{\mathrm{C}}}_\tau}({\tilde{\mathrm{C}}}_\tau) \gets \operatorname{KDE}(\mathscr{F}^{-1}[\tilde{\mathrm{C}}_\kappa], ke, bw)$ 
            \State $\Omega [{\tilde{\mathrm{C}}}_\tau, \Upsilon_{{\tilde{\mathrm{C}}}_\tau}({\tilde{\mathrm{C}}}_\tau), \theta]$ \Comment{Particle Filter set}
    
            \State $\Theta \gets \left[\right]$ \Comment{Particles buffer}
    
            \For{$i$ \text{in} $\Omega$}
        
                \If{$\Omega[i] \geq \epsilon$}
            
                    \State $\Theta \gets \Omega[i]$
            
                \EndIf
            \EndFor
            \State $Y \gets \overline{\Theta}$
    
            \State $\lvert T \rvert \gets \lvert T \rvert -1$
    
        \EndWhile

        \State $W \gets W-1$
    \EndWhile
    \State \textbf{Return:} Y
    \end{algorithmic}
\end{algorithm}

\subsubsection{Hyper-parameters}
\label{hyperarameters}

The following model input parameters can be tuned to adapt to various scenarios:
\begin{itemize}
    \item $\epsilon'$ (DBSCAN):
    \begin{itemize}
        \item This refers to the largest allowable distance between two samples, determining whether one sample is considered part of the neighborhood of the other.
    \end{itemize}
    \item min-samples $ms$ (DBSCAN):
    \begin{itemize}
        \item This parameter specifies the minimum number of samples within a neighborhood.
    \end{itemize}
    \item Kernel $ke$ (KDE):
    \begin{itemize}
        \item To be chosen from \{``gaussian'', ``tophat'', ``epanechnikov'', ``exponential'', ``linear'', ``cosine''\}.
    \end{itemize}
    \item bandwidth $bw$ (KDE):
    \begin{itemize}
        \item The bandwidth of the kernel.
    \end{itemize}
    \item $\theta$: the number of particles to be generated by the inspired particle filter.
    \item $\epsilon$: the probability threshold that a particle weight should exceed in order to be selected.
\end{itemize}
In our case we used the following values: $\epsilon'$ = 0.1, $ms = 15$, $ke = \text{``gaussian''}$, $bw = 0.3$, $\theta = 100$, $\epsilon = 0.98$.

\subsection{Lachesis version 1}
\label{lachesis_v2}

\begin{algorithm}[hbt!]
    \footnotesize
    \caption{Lachesis version 1}\label{alg:lachesis_v2}
    \begin{algorithmic}
    \State \textbf{Variables:} $n,dw,t, \eta$ \Comment{$n$ nodes, $dw$ weekday, $t$ time, $\eta$ forecast flag}

    \Require $Q \gets \overline{q}(n,dw,t)$ \Comment{Average of 4-weeks time series grouped by $n,dw,t$}

    \Require $S \gets \overline{s}(n,dw,t)$ \Comment{Average of 1-week time series grouped by $n,dw,t$}

    \Require $\hat{S} \gets \hat{s}(n,dw,t)$ \Comment{Prediction}

    \State $Y \gets \left[\right]$ \Comment{Result}

    \Ensure{$Q \geq 0$}

    \Ensure{$S \geq 0$}

    \Ensure{$\hat{S} \geq 0$}

    \State $\tilde{S} \gets S -  Q$
    \State $\sigma_{Q,S} \gets \operatorname{cov}(Q,S)$

    \State $\phi \gets 1 + \tilde{S}'$ \Comment{$\tilde{S}'$ is the first derivative of $\tilde{S}$}

    \State $\sigma_{\phi} \gets \operatorname{std}(\phi)$

    \State $Y \gets \ceil{\hat{S} + \sqrt{\left \lvert \phi \cdot \dfrac{\sigma_{Q,S}}{1+\sigma_{\phi}} \right \rvert}\hspace{2pt}}$ 

    \State $ \alpha \gets \left \lvert \dfrac{\lvert \lvert S - \hat{S} \rvert \rvert_2 - \min(\hat{S})}{\max(S) - \min(\hat{S})} \right\rvert $
    \vspace{5pt}
    \If{$\eta$ is True}
        
    \State $Y \gets \ceil{Y \cdot \left (1 + \dfrac{1}{\alpha} \right)}$ \Comment{Forecast}
    \Else
        
        \State $Y \gets \ceil{Y \cdot \left (1 + \dfrac{\alpha}{10^{\floor{\log_{10} \lvert \alpha \rvert}+1}}\right)}$  \Comment{Threshold prediction}
    \EndIf
    \State \textbf{Return:} Y
    \end{algorithmic}
\end{algorithm}

The second version of Lachesis requires a set of nodes for which a previous forecast $\hat{S}$ was generated using Lachesis v0 and data from the previous $4$ weeks $Q$, of the $35$-day historical period.
In this second version, the system uses data from the last week $S$ of the $35$-day historical period, having the same date range as $\hat{S}$, to produce the forecast for the following week.

The premise of the model is to refine the time prediction by taking into account the deviation between $S$ and $\hat{S}$, and between $S$ and $Q$.

The coefficient used is the square root of the absolute value of the first derivative $\phi$ of the deviation between $S$ and $Q$, multiplied by the ratio of the covariance $\sigma_{Q,S}$ to the standard deviation $\sigma_{\phi}$ (see Algorithm \ref{alg:lachesis_v2}).

If the prediction flag $\eta$ is set to true, the result is smoothed by $\alpha$, \ie the normalised Euclidean distance between $\hat{S}$ and $S$, otherwise by the ratio $\dfrac{\alpha}{10^{\floor{\log_{10} \lvert \alpha \rvert}+1}}$. 

$\floor{\log_{10} \lvert \alpha \rvert}+1$ is the number of integer digits of $\alpha$.

The details are described in the Algorithm \ref{alg:lachesis_v2}.


\section{Evaluation}
\label{evaluation}

In this section, we benchmark Lachesis v0 and v1 against the models outlined in Section~\ref{related_work} based on the model development pipeline described in Section~\ref{mlops}.
The models are tuned based on an exploratory data analysis of the time series data, capturing the fundamental temporal characteristics of the data, including lags, periodicity, trend.

\subsection{The First Glance}

\begin{figure} [t] 
    \centering
        \includegraphics[width=\columnwidth]{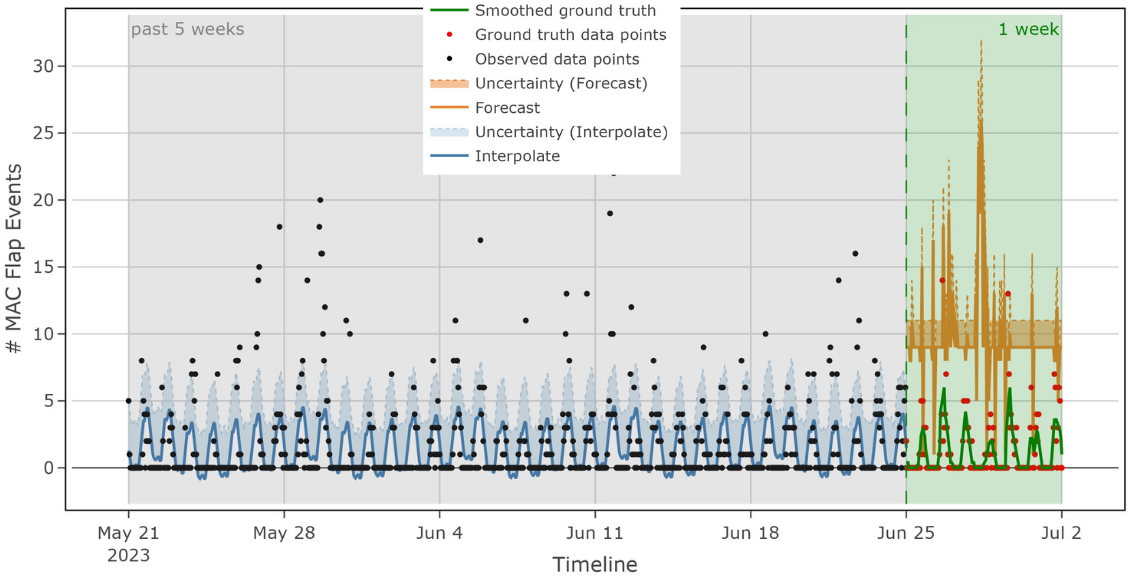}
    \caption{Lachesis v0 prediction}
        \label{fig:forecasting_exe}
        \vskip -.1in
\end{figure}

\begin{figure} [t] 
    \centering
    \includegraphics[width=.8\columnwidth]{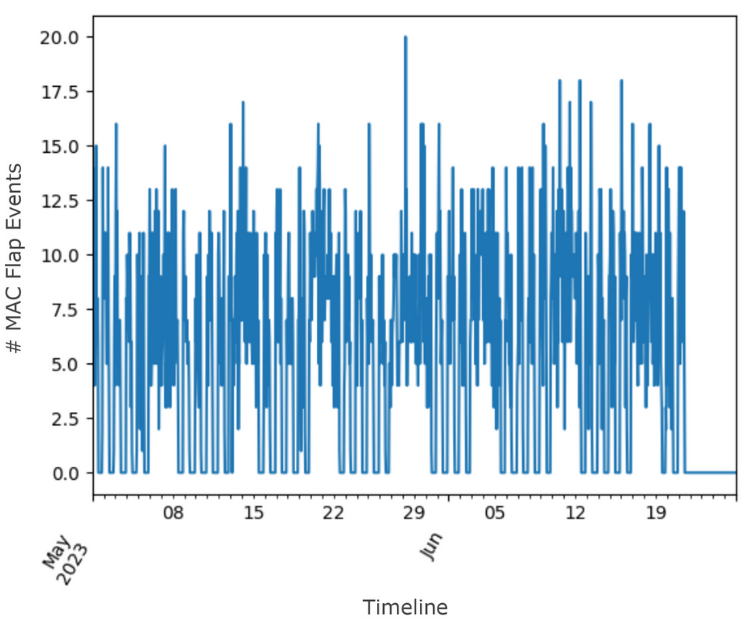}
    \caption{Stationary: an example of a time series classified as high stationary}
    \label{fig:stationarity_full}
    \vskip -.1in
\end{figure}

\begin{table*}[t]
    \centering
    \caption{Performance metrics of a balanced batch of nodes}
    \label{tab:bactch_balanced}
    
    \begin{adjustbox}{width=\textwidth, keepaspectratio}
    \begin{tabular}{c|c|c|c|c|c|c}
        \hline
        \makecell{\textbf{Model}} & \makecell{\textbf{MSE}} & \makecell{\textbf{RMSE}} & \makecell{\textbf{MAE}} & \makecell{\textbf{Avg. Daily Alerted} \\ \textbf{Nodes}} &  \makecell{\textbf{Avg. Daily} \\ \textbf{Alerts}} & \makecell{\textbf{Total Alerts}}  \\
        \hline
        \makecell{arima} & 1769.932±196.916 & 42.027±2.335 & 12.423±0.685 & 56.095±5.288 & 110.714±17.693 & 775±123.851 \\
        \makecell{dlm} & 1769.038±612.300 & 41.622±7.418 & 11.907±2.629 & 71.381±6.671 & 158.571±18.548 & 1110±129.835 \\
        \makecell{lachesis\_v0} & \textbf{4.990±0.227}	& \textbf{2.234±0.051}	& \textbf{1.184±0.018} & 85.667±2.144	& 210.238±9.915	& 1471.667±69.407 \\
        \makecell{lachesis\_v1} & 3430.788±904.687 & 58.23±7.756 & 55.176±7.454 & \textbf{15.048±4.908}	& \textbf{33.333±15.369} & \textbf{233.333±107.584} \\
        \makecell{linear} & 1671.404±318.552	& 40.752±4.003	& 13.624±1.260 & 67.714±5.878	& 150.191±21.976	& 1051.333±153.832 \\
        \makecell{phase1} &15305014±6820400 & 3833.331±957.016	& 563.420±95.862 & 21.857±10.933	& 33.524±19.873	& 234.667±139.113 \\
        \makecell{prophet} & 2353.651±886.237	& 47.953±9.014	& 14.039±2.876 & 65.476±6.118	& 135.524±16.993	& 948.667±118.951 \\
        \makecell{quadratic} & 2018.9±93.3	& 44.924±1.044	& 14.43±1.273 & 66.905±7.593	& 140.619±20.991	& 984.333±146.937 \\
        \makecell{sarima} & 4183.916±706.578	& 64.535±5.359 & 16.669±2.290 & 33.714±3.714	& 60.762±5.915	& 425.333±41.405 \\
        \hline
    \end{tabular}
    \end{adjustbox}

    \vspace{10pt}
    
    \begin{adjustbox}{width=1.4\columnwidth, keepaspectratio}
    \begin{tabular}{c|c|c|c|c}
        \hline
        \textbf{Model} & \makecell{\textbf{t\_train (s) / 1k} \\ \textbf{data points}} & \makecell{\textbf{t\_infer (s) / 1k} \\ \textbf{data points}} & \makecell{\textbf{t\_train (s) / 1k} \\ \textbf{nodes}} & \makecell{\textbf{t\_infer (s) / 1k} \\ \textbf{nodes}} \\
        \hline
        \makecell{arima} & 0.378±0.230	& 0.181±0.099	& 317.738±193.736	& 30.523±16.754 \\
        \makecell{dlm} & 0.685±0.21	& 1.18±0.696	& 576.1±176.809	& 199.337±117.699 \\
        \makecell{lachesis\_v0} & 0.149±0.03	& 0.866±0.138	& 7508.336±1511.362	& 8734.203±1394.595 \\
        \makecell{lachesis\_v1} & 0.3995±0.107	& \textbf{0.0005±0.0002}	& 20132.882±5398.693	& 5.1030±1.98\\
        \makecell{linear} & 0.0057±0.0031	& 0.03±0.015	& 4.8±2.567	& 5.065±2.48 \\
        \makecell{phase1} &\textbf{0.0025±0.0010}	& 0.043±0.013	& \textbf{2.063±0.81}	& 7.297±2.126 \\
        \makecell{prophet} & 0.27±0.06	& 0.756±0.229	& 228.916±50.335	& 127.78±38.72 \\
        \makecell{quadratic} & 0.0055±0.0025	&0.0298±0.0092	&4.654±2.08&	\textbf{5.033±1.554} \\
        \makecell{sarima} & 3.669±1.704	& 0.274±0.168	& 3085.996±1433.175	& 46.365±28.392 \\
        \hline
    \end{tabular}
    \end{adjustbox}
    
    \vspace{10pt}

    \begin{adjustbox}{width=1.4\columnwidth, keepaspectratio}
    \begin{tabular}{c|c|c|c|c}
        \hline
        \makecell{\textbf{Model}} & \makecell{\textbf{TP}} & \makecell{\textbf{FP}} & \makecell{\textbf{TN}} & \makecell{\textbf{FN}} \\
        \hline
    \makecell{arima} & 131.333±24.583 & 643.667±99.962 & 19366.667±187.431 & 6602.667±2978.339 \\
    \makecell{dlm} & \textbf{185.667±22.008} & 924.333±109.441 & 18914±337.23 & \textbf{6548.33±2949.41} \\
    \makecell{lachesis\_v0} & 166±8.19 & 1305.667±64.933 & 14297±576.27 & 6568±2961.09 \\
    \makecell{lachesis\_v1} & 109±42.93 & \textbf{124.33±73.078} & 20154.667±240.80 & 6625±3008.168 \\
    \makecell{linear} & 180.333±21.008 & 871±135.193 & 18823.667±296.659 & 6553.667±2970.17 \\
    \makecell{phase1} & 61.333±11.930 & 173.333±135.711 & \textbf{20293.667±204.962} & 6672.667±2959.571 \\
    \makecell{prophet} & 160.667±14.364 & 788±115.5 & 19118.33±287.77 & 6573.33±2952.52 \\
    \makecell{quadratic} & 177.33±23.71 & 807±126.69 & 18951±424.72 & 6556.67±2953.11 \\
    \makecell{sarima} & 97.33±3.79 & 328±37.8 & 19986±267.1 & 6636.67±2969.4 \\
    \hline
\end{tabular}
\end{adjustbox}

\vspace{10pt}

    \begin{adjustbox}{width=1.5\columnwidth, keepaspectratio}
    \begin{tabular}{c|c|c|c|c|c}
        \hline
        \makecell{\textbf{Model}} & \makecell{\textbf{accuracy}} & \makecell{\textbf{precision}} & \makecell{\textbf{recall}} & \makecell{\textbf{specificity}} 
        & \makecell{\textbf{balanced accuracy}}\\
        \hline
    \makecell{arima} & 0.734±0.078 & 0.169±0.008 & 0.023±0.012 & 0.968±0.005 & 0.495±0.004\\
    \makecell{dlm} & 0.724±0.081 & 0.167±0.007 & 0.0305±0.0110 & 0.953±0.006 & 0.492±0.006\\
    \makecell{lachesis\_v0} & 0.654±0.0901 & 0.113±0.005 & 0.028±0.011 & 0.916±0.004 & 0.472±0.006\\
    \makecell{lachesis\_v1} & \textbf{0.7556±0.0819} & \textbf{0.4786±0.1171} & 0.020±0.014 & \textbf{0.9939±0.0035} & \textbf{0.5070±0.0056}\\
    \makecell{linear} & 0.7245±0.0762 & 0.172±0.011 & \textbf{0.0305±0.0143} & 0.956±0.007 & 0.4931±0.0043\\
    \makecell{phase1} & 0.7535±0.077 & 0.3096±0.1365 & 0.010±0.004 & 0.9916±0.0066 & 0.500±0.001\\
    \makecell{prophet} & 0.729±0.079 & 0.171±0.021 & 0.0264±0.0090 & 0.96±0.006 & 0.493±0.004\\
    \makecell{quadratic} & 0.73±0.08 & 0.18±0.01 & 0.029±0.012 & 0.959±0.007 & 0.494±0.005\\
    \makecell{sarima} & 0.748±0.080 & 0.2297±0.0141 & 0.016±0.007 & 0.984±0.002 & 0.5±0.003\\
    \hline
\end{tabular}
\end{adjustbox}
    
\end{table*}

Figure~\ref{fig:forecasting_exe} depicts an example of the prediction made by Lachesis v0 on a single node.
The smoothed ground-truth trace functions as a reference for determining the precision of the prognosis.
Since networking issues rarely happen, we create a balanced batch of nodes with and without networking issues to thoroughly benchmark various models in Section~\ref{sec:balance_batch}.
Then in Section~\ref{sect:clust_batch}, we consider node batches clustered based on various temporal characteristics, namely stationarity, periodicity, and volatility.

\subsection{A Balanced Batch of Nodes}
\label{sec:balance_batch}

\begin{figure} [t]
    \centering
\includegraphics[width=.8\columnwidth]{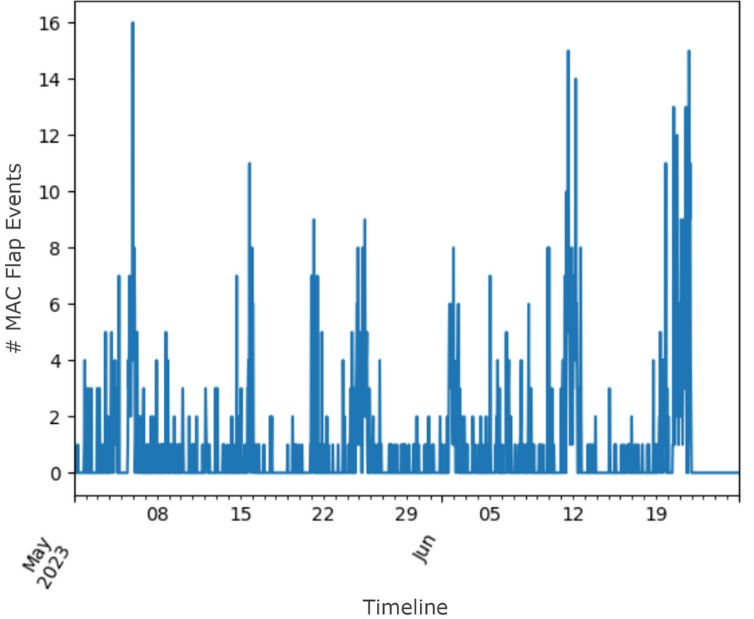}
\caption{Volatility: an example of a time series classified as highly volatile}
\label{fig:volatility_high}
    \vskip -.1in
\end{figure}

We sampled $540$ nodes to make predictions with the inference period spanned over $3$ weeks, from ``2023-06-25" to ``2023-07-16".
A balanced set of nodes was utilised, comprising of $138$ nodes, of which $69$ nodes experienced actual network issues, gathered through user feedback, and $69$ nodes with no known history of network problems.
The previous $5$ weeks were used as the training set.
In Lachesis v1, the fifth week's historical data is used as the prediction basis, building on Lachesis v0's results.
The results are shown in Table~\ref{tab:bactch_balanced}.

The findings indicated improved precision on the Lachesis v0 regression metrics and the number of triggered alerts.
However, with respect to the confusion metrics, there is a small reduction in the Lachesis v0 model's accuracy in comparison to DLM on true positives and false negatives.
Conversely, Lachesis v1 shows the lowest outcomes in false positives prediction, surpassing all other models. 
As for true negatives, Lachesis v1 slightly surpasses phase $1$.
Lachesis v1 illustrates the highest precision and accuracy.

\subsection{Node Batches Clustered on Temporal Characteristics}
\label{sect:clust_batch}
An initial data preparation phase is required to cluster nodes based on time series properties, namely stationarity, periodicity and volatility.

\subsubsection{Stationarity}
\label{stationarity}

The stationarity is determined based on both the Kwiatkowski-Phillips-Schmidt-Shin (KPSS) test \cite{KWIATKOWSKI1992159} and the mean-variance deviation test.
High stationarity is determined following a total agreement of both the tests.
Figure~\ref{fig:stationarity_full} shows an example of node with a high stationary time series.

\subsubsection{Volatility}
\label{volatility}

Let $H$ be a single node time series, volatility is evaluated as follows
\begin{equation}
\nu = \operatorname{std}\left(\dfrac{d(\ln[H+1])}{dt}\right) \cdot \sqrt{\operatorname{card}\left(\dfrac{d(\ln[H+1])}{dt}\right)}
\end{equation}
Clusters are formed using a K-means clustering algorithm on $\nu$ values and nodes are labelled as \textit{high}, \textit{medium} and \textit{low}.
Figure~\ref{fig:volatility_high} shows an example of node experiencing a \textit{high} volatility time series.

\subsubsection{Periodicity}

\begin{figure} [t] 
    \centering
    \includegraphics[width=.8\columnwidth]{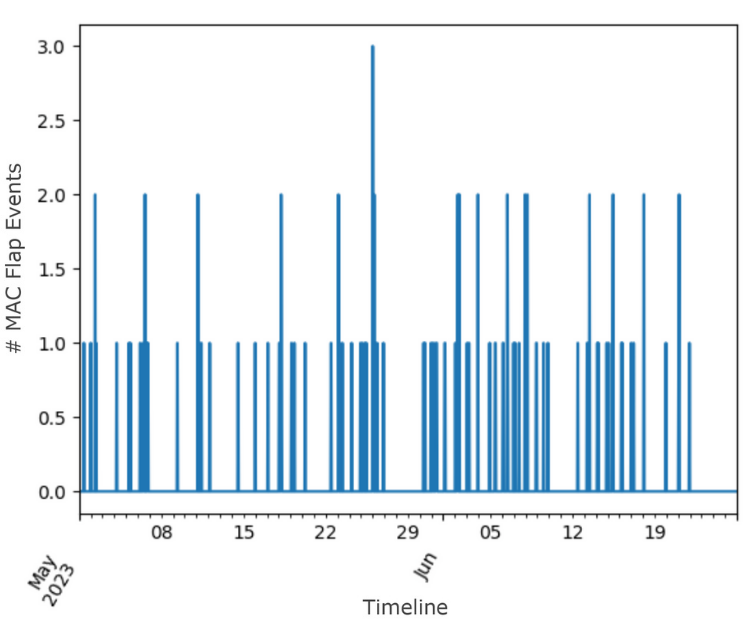}
    \caption{Periodicity: an example of a time series classified as highly periodic}
    \label{fig:periodicity_high}
    \vskip -.1in
\end{figure}

A customized formula has been designed to assess the recurrence of MAC-flap events on the same day of the week and time of day. This is to evaluate the underlying assumption of the Lachesis model, that anomalies due to non-malicious causes recur consistently on the same day of the week and time of day, whereas malicious anomalies are associated with sporadic occurrences.
\begin{equation}
    \mathrm{R}_{dw, t_\tau} = \dfrac{1}{1+2\sigma'\mathbb{P}_{dw, t_\tau}(\mathrm{\textit{mode}})} \quad \forall dw, t_\tau \in H_\tau
\end{equation}
with
\begin{equation}
    \mathbb{P}_{dw, t_\tau}(\textit{mode}) = \dfrac{\max\limits_{\omicron \in \mathrm{O}} \left(f_\omicron\right)}{\sum_{\omicron \in \mathrm{O}} f_\omicron}
\end{equation}
such that $\mathrm{O}$ is the set containing the size of events, in terms of number of alerts, and $f_\omicron$ is the relative frequency of occurence.
Figure \ref{fig:periodicity_high} shows an example of node experiencing a highly periodic time series.

\begin{table*}[t]
    \centering
    \caption{Performance metrics of a batch of nodes with high periodicity}
    \label{tab:period_high}
    
    \begin{adjustbox}{width=\textwidth, keepaspectratio}
    \begin{tabular}{c|c|c|c|c|c|c|c|c|c}
        \hline
        \makecell{\textbf{Model}} & \makecell{\textbf{MSE}} & \makecell{\textbf{RMSE}} & \makecell{\textbf{MAE}} & \makecell{\textbf{Avg. Daily Alerted} \\ \textbf{Nodes}} &  \makecell{\textbf{Avg. Daily} \\ \textbf{Alerts}}  & \makecell{\textbf{Avg. Alert} \\ \textbf{duration (min.)}} & \makecell{\textbf{Std. Alert} \\ \textbf{duration (min.)}} & \makecell{\textbf{Avg. Alerts} \\ \textbf{per hour}} & \makecell{\textbf{Std. Alerts} \\ \textbf{per hour}} \\
        \hline
        \makecell{arima} & 181.675	&8.116	&4.330	&53.3	&111.3   & 110.76&	100.127	&4.726&	2.708\\
        \makecell{dlm} & 212.795	&9.207&	5.590&	64.2&	143.5  & 108.68	&105.475&	6.021&	2.869 \\
        \makecell{lachesis\_v0} & \textbf{1.441}	&\textbf{1.143}	&\textbf{1.003}	&76.3&	188.5   & 217.37&	311.608	&7.854	&3.491\\
        \makecell{lachesis\_v1} & 2384.963	&48.532	&48.052&	\textbf{8.8}	&\textbf{16.3}  & 101.44	&\textbf{69.968}	&\textbf{1.468}&	\textbf{0.889}\\
        \makecell{linear} & 176.812	&8.492&	5.182	&64.0	&151.3  & 121.16	&123.672&	6.369	&3.020 \\
        \makecell{phase1} &9280.777&	42.813&	19.870&	30.2&	49.5   &\textbf{96.98}	&77.192	&2.753	&1.656 \\
        \makecell{prophet} & 200.930	&8.954	&5.729	&59.2	&128.8  & 109.34	&95.463&	5.496&	2.708\\
        \makecell{quadratic} & 217.408	&9.319	&6.372	&61.7	&136.0  & 114.34	&111.667&	5.693&	3.068 \\
        \makecell{sarima} & 303.269	&9.430&	5.714&	33.0&	56.8  & 103.73	&96.163&2.684&	1.635 \\
        \hline
    \end{tabular}
    \end{adjustbox}
    
    \vspace{10pt}

    \begin{adjustbox}{width=.7\textwidth, keepaspectratio}
    \begin{tabular}{c|c|c|c|c|c|c|c|c|c}
        \hline
        \makecell{\textbf{Model}} & \makecell{\textbf{TP}} & \makecell{\textbf{FP}} & \makecell{\textbf{TN}} & \makecell{\textbf{FN}} & \makecell{\textbf{accuracy}} & \makecell{\textbf{precision}} & \makecell{\textbf{recall}} & \makecell{\textbf{specificity}}  & \makecell{\textbf{balanced accuracy}}\\
        \hline
        \makecell{arima} & 88	&580	&17153	&152	&0.959&	0.132	&0.367	&0.967&	0.667\\
        \makecell{dlm} & 97	&764&	16835&	\textbf{122}	&0.950	&0.113&	\textbf{0.443}&	0.957	&\textbf{0.700}\\
        \makecell{lachesis\_v0} & 69&	1062	&14341	&306&	0.913	&0.061	&0.184	&0.931&	0.558 \\
        \makecell{lachesis\_v1} & 44	&\textbf{54}	&\textbf{18198}	&284	&\textbf{0.982}&	\textbf{0.449}&	0.134	&\textbf{0.997}	&0.566\\
        \makecell{linear} & \textbf{101}&	807&	16562&	139	&0.946&	0.111&	0.421	&0.954	&0.687 \\
        \makecell{phase1} &33	&264	&17822	&538	&0.957&	0.111&	0.058	&0.985&	0.522\\
        \makecell{prophet} & 76	&697	&16955	&171	&0.952	&0.098&	0.308&	0.961&	0.634 \\
        \makecell{quadratic} & 92	&724	&16829	&147	&0.951&	0.113&	0.385	&0.959	&0.672 \\
        \makecell{sarima} & 54	&287	&17776	&268	&0.970	&0.158	&0.168	&0.984	&0.576 \\
        \hline
    \end{tabular}
    \end{adjustbox}
    
\end{table*}

\begin{table*}[t]
    \centering
    \caption{Performance metrics of a batch of nodes with full stationary}
    \label{tab:station_full}
    
    \begin{adjustbox}{width=\textwidth, keepaspectratio}
    \begin{tabular}{c|c|c|c|c|c|c|c|c|c}
        \hline
        \makecell{\textbf{Model}} & \makecell{\textbf{MSE}} & \makecell{\textbf{RMSE}} & \makecell{\textbf{MAE}} & \makecell{\textbf{Avg. Daily Alerted} \\ \textbf{Nodes}} &  \makecell{\textbf{Avg. Daily} \\ \textbf{Alerts}}  & \makecell{\textbf{Avg. Alert} \\ \textbf{duration (min.)}} & \makecell{\textbf{Std. Alert} \\ \textbf{duration (min.)}} & \makecell{\textbf{Avg. Alerts} \\ \textbf{per hour}} & \makecell{\textbf{Std. Alerts} \\ \textbf{per hour}} \\
        \hline
        \makecell{arima} & 3103.485	&28.589	&20.025	&38.667	&86.167 & \textbf{102.180}	&\textbf{83.891}	&3.743&	1.874 \\
        \makecell{dlm} & 4180.125	&31.871	&23.351&	47.333&	111.167 & 105.361	&100.870	&4.663&	2.320 \\
        \makecell{lachesis\_v0} & \textbf{8.148}&	\textbf{1.857}	&\textbf{1.339}&	54.167&	142.667  & 307.934	&437.898&	5.975&	3.029 \\
        \makecell{lachesis\_v1} & 2630.949	&49.714	&47.448	&\textbf{19.833}	&50.833  &144.803&	174.928&	2.584&	1.420 \\
        \makecell{linear} & 3074.893	&29.547	&22.516	&47	&120.333 & 120.445	&118.650&	5.159&	2.447\\
        \makecell{phase1} &13406608.127&	1003.923&	789.627	&22.667&	\textbf{36.500} &109.924&	94.509&\textbf{	2.229}	&\textbf{1.316}\\
        \makecell{prophet} & 5826.676	&35.251	&26.949&	45&	101.500  & 110.972&	103.088&	4.374&	2.071 \\
        \makecell{quadratic} & 3484.138	&30.256	&23.721	&45.667&	105.667  & 110.347&	106.205&	4.584&	2.192\\
        \makecell{sarima} & 6604.831	&38.787	&29.096	&25.333	&45  & 113.333	&104.820&	2.390	&1.348 \\
        \hline
    \end{tabular}
    \end{adjustbox}
    
    \vspace{10pt}

    \begin{adjustbox}{width=.7\textwidth, keepaspectratio}
    \begin{tabular}{c|c|c|c|c|c|c|c|c|c}
        \hline
        \makecell{\textbf{Model}} & \makecell{\textbf{TP}} & \makecell{\textbf{FP}} & \makecell{\textbf{TN}} & \makecell{\textbf{FN}} & \makecell{\textbf{accuracy}} & \makecell{\textbf{precision}} & \makecell{\textbf{recall}} & \makecell{\textbf{specificity}}  & \makecell{\textbf{balanced accuracy}}\\
        \hline
        \makecell{arima} & 117	&400	&17243	&1530&	0.900&	0.226&	0.071	&0.977	&0.524 \\
        \makecell{dlm} & 136	&531	&16970&	1401	&0.899&	0.204&	0.088	&0.970	&0.529 \\
        \makecell{lachesis\_v0} & 114	&742	&13879	&1245	&0.876	&0.133&	0.084&	0.949&	0.517 \\
        \makecell{lachesis\_v1} &126	&179	&17396	&1101&	\textbf{0.932}&\textbf{0.413}&	0.103	&0.990	&0.546\\
        \makecell{linear} & \textbf{150} &572	&16713	&\textbf{925} & 0.918	&0.208&	\textbf{0.140}	&0.967	&\textbf{0.553}\\
        \makecell{phase1} &46	&\textbf{173}	&\textbf{17616}	&2604	&0.864	&0.210&	0.017&	\textbf{0.990}&	0.504 \\
        \makecell{prophet} & 108	&501&	16983	&1919	&0.876&	0.177&	0.053&	0.971&	0.512 \\
        \makecell{quadratic} & 138	&496	&16983&	1192	&0.910	&0.218	&0.104&	0.972&	0.538 \\
        \makecell{sarima} & 72	&198	&17576&	2179&	0.881	&0.267&	0.032&	0.989&	0.510 \\
        \hline
    \end{tabular}
    \end{adjustbox}
\end{table*}

\begin{table*}[t]
    \centering
    \caption{Performance metrics of a batch of nodes with high volatility}
    \label{tab:volatility_high}
    
    \begin{adjustbox}{width=\textwidth, keepaspectratio}
    \begin{tabular}{c|c|c|c|c|c|c|c|c|c}
        \hline
        \makecell{\textbf{Model}} & \makecell{\textbf{MSE}} & \makecell{\textbf{RMSE}} & \makecell{\textbf{MAE}} & \makecell{\textbf{Avg. Daily Alerted} \\ \textbf{Nodes}} &  \makecell{\textbf{Avg. Daily} \\ \textbf{Alerts}}  & \makecell{\textbf{Avg. Alert} \\ \textbf{duration (min.)}} & \makecell{\textbf{Std. Alert} \\ \textbf{duration (min.)}} & \makecell{\textbf{Avg. Alerts} \\ \textbf{per hour}} & \makecell{\textbf{Std. Alerts} \\ \textbf{per hour}} \\
        \hline
        \makecell{arima} & 8574.378	&56.854	&44.696	&12.5	&27.5   & 154.322	&177.520&	1.823	&0.954\\
        \makecell{dlm} & 12175.386	&65.003	&50.037&	14.5	&31.5   & 118.017	&102.821&	1.765	&0.964 \\
        \makecell{lachesis\_v0} & \textbf{21.970}	&\textbf{3.050}	&\textbf{2.027}	&16.833&	41.333   & 396.687	&536.435&	2.167&	1.411 \\
        \makecell{lachesis\_v1} & 3429.130	&54	&46.379	&9.667	&25.833  & 172.856	&226.995&	1.720&	0.929 \\
        \makecell{linear} & 8657.113	&60.636	&51.350	&13.667	&32   & 159.310	&153.114	&1.883	&1.072 \\
        \makecell{phase1} &45179583.240	&3128.713	&2521.655	&7.167	&12.167   &137.501	&102.494&	1.384	&0.511 \\
        \makecell{prophet} & 17761.725	&76.574	&61.494&	15.167&	31.500  & 139.667&	130.948	&1.759&	1.049 \\
        \makecell{quadratic} & 9752.012	&58.776	&49.066	&14.5	&32   &132.414	&125.119&	1.893	&1.090 \\
        \makecell{sarima} & 19906.523	&86.075	&68.089&	\textbf{6.667}	&\textbf{10.833}  & \textbf{107.337}	&\textbf{78.960}&	\textbf{1.232}&	\textbf{0.407}\\
        \hline
    \end{tabular}
    \end{adjustbox}

    \vspace{10pt}

    \begin{adjustbox}{width=.7\textwidth, keepaspectratio}
    \begin{tabular}{c|c|c|c|c|c|c|c|c|c}
        \hline
        \makecell{\textbf{Model}} & \makecell{\textbf{TP}} & \makecell{\textbf{FP}} & \makecell{\textbf{TN}} & \makecell{\textbf{FN}} & \makecell{\textbf{accuracy}} & \makecell{\textbf{precision}} & \makecell{\textbf{recall}} & \makecell{\textbf{specificity}}  & \makecell{\textbf{balanced accuracy}}\\
        \hline
        \makecell{arima} &33	&132	&17880&	1331&	0.924&	0.200&	0.024	&0.993	&0.508 \\
        \makecell{dlm} & 46	&143&	17942&	1222&	0.929&	0.243&	0.036&	0.992&	0.514 \\
        \makecell{lachesis\_v0} & 56	&192&	16758&	917&	0.938&	0.226&	0.058&	0.989	&0.523\\
        \makecell{lachesis\_v1} & \textbf{59}	&96	&17905&	841	&0.950&	\textbf{0.381}&\textbf{	0.066}&	0.995&	\textbf{0.530} \\
        \makecell{linear} & 47	&145	&17807&	\textbf{765}&	\textbf{0.952}&	0.245&	0.058&	0.992&	0.525 \\
        \makecell{phase1} &21	&52	&18108	&2015	&0.898&	0.288&	0.010&	0.997&	0.504 \\
        \makecell{prophet} & 37	&152&	17864&	1701&	0.906&	0.196&	0.021&	0.992&	0.506 \\
        \makecell{quadratic} & 52	&140&	17899&	999&	0.940&	0.271	&0.049	&0.992&	0.521 \\
        \makecell{sarima} & 22	&\textbf{43}	&\textbf{18176}	&1876	&0.905	&0.338&	0.012&	\textbf{0.998}&	0.505\\
        \hline
    \end{tabular}
    \end{adjustbox}
    
\end{table*}

\subsubsection{Results}

A group of nodes was analysed, with $71$ being classified as having \textit{high} stationarity, $105$ demonstrating \textit{high} periodicity, and $21$ nodes exhibiting \textit{high} volatility.
The inference period lasted for one week, between ``2023-06-25" and ``2023-07-02".

The assessment primarily focused on the regression and alerting accuracy, as the computational complexity over time had been previously evaluated.
Therefore, an expanded version of these metrics is presented in Tables \ref{tab:period_high}-\ref{tab:volatility_high}.

The findings revealed that Lachesis v0 exhibited consistent leadership on the regression metrics across all three clusters, surpassing the performance of the other models. However, v1's primate on the alertness and confusion metrics is only sustained within the periodic cluster, aligning with the Lachesis grounded hypothesis, as illustrated in Table \ref{tab:period_high}.

Under the assumption of complete stationarity, the phase $1$ model exhibited the best overall performance regarding confusion and alerting metrics. Both versions of Lachesis demonstrated strong results compared to the other models, particularly maintaining a leadership position in terms of accuracy and precision metrics, see Table \ref{tab:station_full}.

In the volatility cluster, SARIMA exhibited the best performance in terms of daily alarms as well as the metrics for alarm duration and false positive predictions. Nonetheless, Lachesis v1 outperformed all models in terms of true positive detection and overall user experience related metrics, with Linear Regression coming in second place as shown in Table \ref{tab:volatility_high}.

\section{Conclusion}
\label{conclusion}

In this paper, we proposed a systematic model development pipeline, which embraces a holistic approach that accords equal primacy to technical prowess and user experience metrics.
Through the integration of user-centric feedback systems, our framework facilitates the seamless scaling of various models across diverse clusters of nodes, thereby enhancing evaluative precision and validation accuracy.
This pipeline helped us develop an integrated anomaly detection system that employs a particle-filter-based algorithm named \emph{Lachesis}.
Results highlighted the superior forecasting capabilities of Lachesis v0 compared with existing models in literature, while Lachesis v1 excelled in anomaly detection.
Rigorous experimentation, which included scenarios concerning nodes affected by network issues, and clusters of nodes based on temporal characteristics, demonstrated the effectiveness of the two algorithms.
This paper demonstrated the capacity of the proposed model development pipeline in enabling the efficient and continuous iteration of data-driven products, setting a new benchmark for responsive and user-centric anomaly detection systems.

\bibliographystyle{IEEEtran}
\bibliography{reference}

\end{document}